\documentclass[twocolumn,notitlepage,prb,superscriptaddress,showpacs]{revtex4-1}

\usepackage{hyperref}
\usepackage[usenames,dvipsnames]{color}
\usepackage{amsmath}
\usepackage[english]{babel}
\usepackage{graphicx}
\usepackage{epstopdf}
\usepackage{lineno}
\usepackage{tabularx}
\usepackage{multirow}

\usepackage[utf8]{inputenc}
\usepackage{graphicx}
\graphicspath{ {./images/} }

\begin{document}

\title{Fluctuation diagnostics of the finite temperature quasi-antiferromagnetic regime of the 2D Hubbard model.}
\author{Behnam Arzhang}
\affiliation{Department of Physics and Physical Oceanography, Memorial University of Newfoundland, St. John's, Newfoundland \& Labrador A1B 3X7, Canada} 
\author{A. E. Antipov}
\affiliation{Microsoft Quantum, Microsoft Station Q, University of California, Santa Barbara, California 93106-6105 USA}
\author{J. P. F. LeBlanc}
\email{jleblanc@mun.ca}
\affiliation{Department of Physics and Physical Oceanography, Memorial University of Newfoundland, St. John's, Newfoundland \& Labrador A1B 3X7, Canada} 

\date{\today}

\begin{abstract}
We study the finite temperature Fermi-liquid to non-Fermi-liquid crossover in the 2D Hubbard model for a range of dopings using the self-consistent ladder dual fermion method.
 We consider relatively high temperatures where we identify a suppression of the density of states near the Fermi level caused by a quasi-antiferromagnetic behaviour that is itself characterized by a long, but finite, correlation length scale.  We perform fluctuation diagnostics to decompose the single-particle self energy into scattering $q$-vector and bosonic frequency contributions.  Within this framework we find that the key contributions to the single-particle self energy that give non-Fermi-liquid character, even at weak coupling, are caused by relatively sharp $q=(\pi,\pi)$ spin fluctuations, while the decomposition in the bosonic frequency channel shows a complicated dependence on the relative strengths of zero, positive and negative frequency contributions.  Finally, variation in density suggests that the tendency towards non-Fermi-liquid behavior is not substantially different for electron or hole doped systems.
\end{abstract}

\maketitle

\section{Introduction \label{sec:introduction}}
Understanding the $T-U$ phase diagram of the 2D Hubbard model on a square lattice has been a key problem  in the field of strongly correlated electron systems that has driven numerical methods development.
While the ground state of the half-filled model is expected to be an antiferromagnetic band insulator, the finite temperature phase diagram of both the half-filled and particle-hole asymmetric cases has been a topic of controversy.\cite{schafer:2015}
Most recently, Simkovic et al. \cite{simkovic:2019, simkovic2017determinant} has shown that for at least one metric of Fermi-liquid behavior, the single-particle self energy\cite{park:2008},  the Fermi-liquid (FL) to non-fermi-liquid (nFL) crossover has an exponential onset in the T-U phase diagram consistent with a N\'eel temperature mean-field ansatz $T_N\propto e^{-b/\sqrt{U}}$. 
This is in contrast to what has become the standard picture of the 2D Hubbard model phase diagram,\cite{park:2008} that there should be some set of critical values $U_{c1}$ and $U_{c2}$ across which properties such as double occupancy show hysteresis, an indication of a first order phase transition to an interaction driven Mott-insulator. 
Calculations based on cluster extensions of dynamical mean-field theory (DCA, cDMFT), typically assume the system to be paramagnetic and suppress antiferromagnetic correlations during the self-consistency.  This is done under the very reasonable presumption that in the limit of infinite cluster size the Mermin-Wagner theorem would prevent a  second order phase transition to an antiferromagnetic state.  
What was not expected is that antiferromagnetic spin-correlations with a finite correlation length scale could be sufficient to lead to a nFL state.  
It appears that small-cluster cDMFT or DCA calculations in the weak coupling regime and at low-temperatures incorrectly represent the correlation length-scale due to a truncation of the system size.\cite{georges:1996, leblanc:2013}
Thus far there has been much conjecture regarding the role of these finite-ranged AFM fluctuations.\cite{schafer:2015}
The question of precisely how these spin fluctuations cause a suppression of states near the Fermi level has not been answered.

In order to address this problem, we employ a recent development due to  Gunnarsson et al.\cite{gunnarsson:2015}. Their primary insight, called fluctuation diagnostics, allows one to make use of two-particle vertex methods to decompose the single-particle self-energy into various basis representations and scattering channels.   The original work on 8-site DCA data suggested that contributions to the self-energy giving rise to a pseudogap were dominated by $q=(\pi,\pi)$ spin fluctuations. Further, they showed unequivocally that the physical process resulting in pseudogap physics in the 2D Hubbard model is best represented by spin excitations rather than in the charge or particle-particle channels, an observation supported by diagrammatic application of the same fluctuation diagnostics procedure.\cite{wu:2017} 
To provide a complete picture of scattering vector contributions one requires a numerical method with high momentum resolution that can capture correlations both at short and long distance.  For this reason, we employ the method of dual fermions\cite{rubtsov:2008} (DF) within the self-consistent ladder approximation.\cite{opendf}   As an extension of the DMFT, the DF method has the advantage that it correctly captures both the short range correlations (through DMFT) and the long range correlations (via the duality transformation) and also can generate both single and two-particle quantities on an arbitrarily fine momentum grid with a roughly linear expense in resolution.  

In this work, we utilize the power of the fluctuation diagnostics method within the DF framework to examine precisely how excitations in the spin channel cause suppression of states near the Fermi level ultimately resulting in a fermi-liquid (FL) to non-Fermi-liquid (nFL) crossover in the 2D Hubbard model.  We extract the imaginary part of the self energy and examine $\Delta \Sigma = Im\Sigma (i\omega_0) - Im\Sigma (i\omega_1)$ as a metric of FL/nFL behaviour.\cite{park:2008, simkovic:2019}  We examine weak, intermediate and strong interaction strengths at fixed temperature with nominal $t^\prime$ and subsequently dope the system to regain FL character. \cite{wu:2017,schafer:2015}  We then study in detail the low temperature weak coupling case, in order to first verify the existence of the nFL crossover at finite T for the half filled model\cite{simkovic:2019} and then comment precisely on the role of spin fluctuations in producing that specific nFL behaviour. We find that non-Fermi-liquid behavior arises almost entirely from $(\pi,\pi)$ spin-correlations, and as both a function of doping and temperature coincides with an increase in correlation length scale.  Further, we note that the extent of non-fermi liquid behavior is not strongly dependent upon electron or hole doping, but that the $q$-vector decompositions appear quite distinct.

In section~\ref{sec:methods} we overview the applied numerical tools and theory, section~\ref{sec:results} presents our results and section~\ref{sec:conclusions} concludes our discussion.

\section{Theory and Methods \label{sec:methods}}

We restrict our study to the single-band Hubbard model on a 2D square lattice with nearest and next-nearest neighbour tight binding dispersion.\cite{benchmarks}  The Hamiltonian is given by
\begin{equation}
H = \sum_{k,\sigma} \left(\epsilon_{k} -\mu\right)c_{k\sigma}^\dagger c_{k\sigma}+U\sum_i n_{i\uparrow}n_{i\downarrow},
\label{H}
\end{equation}
where $\mu$ is the chemical potential, $k$ momentum, $i$ labels of sites in real-space, $U$ the interaction, and the dispersion is given by
$\epsilon_k=-2t\left[\cos(k_x)+\cos(k_y)\right]-4t^\prime \cos(k_x)\cos(k_y)$ where $t$ and $t^\prime$ are the nearest and next-nearest-neighbor hopping terms respectively.

\subsection{Ladder Dual-Fermions}
In the dual-fermion formalism\cite{rubtsov:2008,opendf} one replaces the lattice problem with a lattice of coupled Anderson impurity problems resulting in an action of the form
\begin{equation}
    S[f,f^*]= \sum_{\omega k \sigma}\text{g}_\omega^{-2} \left((\Delta_\omega - \epsilon_k)^{-1}
    +\text{g}_\omega\right) f^*_{\omega k \sigma} f_{\omega
    k \sigma} + \sum_i V_i,
\end{equation}
with $V_i\equiv V[f^*_i,f_i]$, and where $\text{g}_w$ is the momentum independent Green's function of the Anderson impurity problem, and $\Delta_\omega$ is the hybridization function between the impurity and the bath.\cite{georges:1996,rubtsov:2008}  Unlike Eq.~(\ref{H}) which depends on electronic creation and annihilation operators, the dual-fermion action now depends on $f$ and $f^*$ which are dual operators obtained via a Hubbard-Stratonovich transformation.  

In the dual space the interactions have become local (the sum over sites $i$ above) and are collected into the function $V_i$.  
The mapping is formally exact and is simply an expansion as a function of  $\text{g}_w$ and the $n$-leg impurity vertex functions, $\gamma^{(n)}$. 
Correctly representing $V_i$ remains problematic due to the complexity of higher-leg vertex functions\cite{ribic:2017} and so we truncate the vertex at the level of 4-leg operators, $\gamma^{(4)}$.
In this sense we have an approximate method to solve Hubbard-like models.   Where one would need to sum all diagrammatic contributions at all orders in $\gamma^{(4)}$.\cite{iskakov:2016,gukelberger:2017, rohringer:revmodphys} Once the dual self energy, $\tilde{\Sigma}(k,\omega)$ is obtained, the lattice self energy is given by
\begin{equation}
\Sigma(k,\omega)=\frac{\tilde{\Sigma}(k,\omega)}{1-\text{g}_{\omega}\tilde{\Sigma}(k,\omega) }+\Sigma^{DMFT}(\omega).
\label{eqn:dualselfenergy}
\end{equation}

\emph{In this work, we employ the self consistent ladder-DF approximation in which only a small subset of diagrams is included, namely a set of vertex ladder diagrams in the charge and spin channels. }
The result is then iterated self consistently until convergence.
To perform these tasks we utilize our open source code \begin{tt}openDF\end{tt}\cite{Antipov15} which produces the self energy $\Sigma(k,i\omega_n)$, the associated Green's functions $G(k,i\omega_n)$ and spin susceptibilities $\chi_s(Q,i\omega_m)$ on the Matsubara axis as well as the self energy, discussed in the next section, decomposed into scattering $q$-vector and bosonic frequency $\Omega$ contributions required for the fluctuation diagnostics analysis. 
Within the dual fermion method single particle self energies are treated self consistently  and this has resulted in typically excellent results.\cite{gukelberger:2017,benchmarks} The status of two particle susceptibilities from DF is known to be approximate and expected to maintain only qualitative correctness.\cite{leblanc:2019}
Input DMFT solutions are obtained using a continuous time auxilliary field method (CT-AUX).\cite{gull:2011}  When necessary we employ analytic continuation using the maximum entropy inversion code MaxEnt implemented in Ref.\onlinecite{maxent}, and crosscheck against other codes that are freely available.\cite{Gaenko17,wallerberger:2018,triqs}

\subsection{Fluctuation Diagnostics from DF}

The single particle self-energy can be constructed from the full two-particle vertex function in the spin basis, $F_{sp}$,\cite{Rohringer12} and the fully interacting lattice Green's function $g(k,\omega)$ (a function of both momentum and frequency) via

\begin{equation}\label{eqn:fluct}
     \Sigma(k,\omega)= \frac{Un}{2}+ \frac{U}{\beta^2 N} \sum \limits_{\bar{\omega}^\prime ,\bar{\Omega}} F_{sp}^{\bar{\omega},\bar{\omega}^\prime,\bar{\Omega}} g(\bar{\omega}^\prime)g(\bar{\omega}^\prime +\bar{\Omega}) g(\bar{\omega}+\bar{\Omega})
\end{equation}
where we use a multi-index for momentum-frequency pairs, $\bar{\omega}=(k,\omega)$, $\bar{\omega}^\prime=(k^\prime,\omega^\prime)$, $\bar{\Omega}=(q,\Omega)$ and we take frequencies $\omega$ and $\omega^\prime$ to represent fermionic matsubara frequencies while $\Omega$ represents a bosonic Matsubara frequency.
As discussed in Ref.~\onlinecite{gunnarsson:2015} the basis choice of the full vertex has no impact on the self energy after summation over all internal indices of the vertex.  We study only the spin channel, since past work has suggested that for the single band model the charge and particle-particle channels are less structured and we omit the Hartree shift throughout.

We then probe the partial summation of the second term in Eq.~\eqref{eqn:fluct} for the spin channel.  This means we decompose the self energy into a more general $q$ and $\Omega$ dependent object
\begin{equation}
    \Sigma(\bar{\omega})= \sum\limits_{\bar{\Omega},\bar{\omega}^\prime} \Sigma_{sp}(\bar{\omega},\bar{\omega}^\prime, \bar{\Omega}).
\end{equation}
To assist with notation we drop the spin, `sp', subscript and suppress the explicit momentum $k$ argument in favour of a subscript to represent the self energy at the so-called nodal (N, $k=(\pi/2,\pi/2)$) and antinodal (AN, $k=(\pi,\pi)$) momenta.

One challenge with this form of decomposition is the high dimensionality of the fully decomposed object $\Sigma(\bar{\omega},\bar{\omega}^\prime, \bar{\Omega})$.  The original work by Gunnarsson et al.\cite{gunnarsson:2015} primarily studied the decomposition of the self energy for the zeroeth fermionic matsubara frequency $Im\Sigma_k(i\omega_0)$ in $q_x$ and $q_y$ for 8 points in momentum space due to the small DCA cluster employed and made some comment on the role of positive bosonic matsubara frequencies.  We wish to study the complete set of all deconstructions, and to assist with this we introduce a shorthand for partial summations of that object given by
\begin{equation}\label{eqn:sumnotation}
    \Sigma_k^{(x)}= \sum\limits_{x, \bar{\omega}^\prime} \Sigma(\bar{\omega},\bar{\omega}^\prime, \bar{\Omega}).
\end{equation}

We will present partial summations over: scattering momenta, $x=q$ ;  positive, negative or all bosonic frequencies, $x= +\Omega, -\Omega or \Omega$ respectively; or over combinations of variables such as $x=(+\Omega, q)$ which represents summation over positive bosonic frequencies and all q-vectors. Also, we always sum explicitly over the internal primed fermionic elements $\bar{\omega}^\prime$ in order to reduce our deconstructed dimensionality which is convenient since $\bar{\omega}^\prime$ does not appear, for our notation, in either single particle self energies or two particle susceptibilities.

Throughout we restrict our solutions of the two particle Green's function and associated vertex of the impurity problem to a truncated set of fermionic and bosonic frequencies, $\Omega= -32 \to 32$ and $\omega,\omega^\prime = -64 \to 63$ inclusive.  We have verified that this is a sufficiently large truncation to accurately reconstruct the  DMFT and DF self energies via Eq.~\eqref{eqn:fluct}.
Further, we will examine the metric for FL/nFL behaviour $\Delta \Sigma_k= Im\Sigma_k(i\omega_0)-Im\Sigma_k(i\omega_1)$ for each decomposed  or partially summed channel, which we will abbreviate as $\Delta \Sigma _{k}^{(x)}$ as per Eq.~(\ref{eqn:sumnotation}).

\begin{figure}
\centering
    \includegraphics[width=0.95\linewidth]{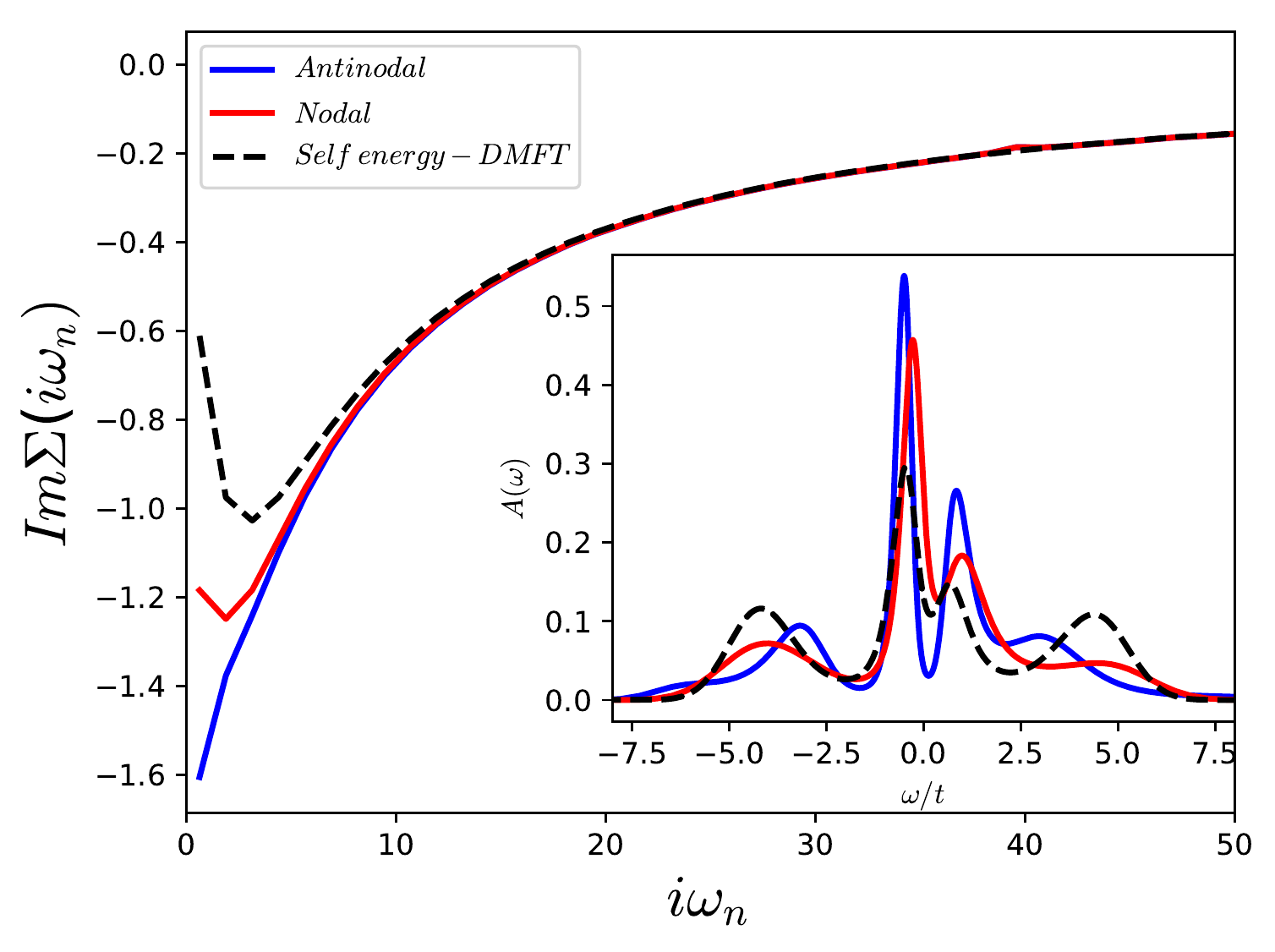}
        \caption{\label{fig:selfedmft} Imaginary part of the self energy as a function of Matsubara index for the DF nodal (N) and antinodal (AN) results as well as the DMFT result on which DF is based for parameters $U/t=5.6$, $\beta t=5$, $t^\prime /t=-0.3$ and $\mu=0$.  Inset: Analytic continuation result for the spectral function, $A(\omega)$ for real frequency $\omega$.}
\end{figure}

\section{Results  \label{sec:results}}

\subsection{Doping Dependent Crossover}
As an illustrative example, in Fig.~\ref{fig:selfedmft} we present DF results at $U/t=5.6$, $t^\prime/t=-0.3$, $\mu=0$, $\beta t=5$ for the imaginary part of the self energy. The DMFT result (dashed-black) shows a tendency towards FL behaviour where the first Matsubara frequency has a higher value than the second ($\Delta \Sigma > 0$).  Shown in red and blue are results at the nodal and antinodal momenta respectively from the DF calculation which provides momentum dependence to the self-energy. 
These results are in agreement with  DiagMC and DCA calculations from  Ref.~\onlinecite{wu:2017} for similar parameters.  We observe for these parameters the shift from FL to partial nFL behaviour indicated by the negative value of $\Delta \Sigma$ at the antinodal point, while the nodal point remains with $\Delta \Sigma >0$.
This behavior is often referred to as the pseudogap phenomenon.\cite{gull:2009, ferrero:2009, yrzreview:2012, leblanc:2011} 

It is worth noting that since these results for $\beta t=5$ are at relatively high temperature, finding $\Delta \Sigma <0$ may not be indicative of a fully gapped state.  
To illustrate this, we perform analytic continuation\cite{maxent} for the local-DMFT Green's function, and the Green's function for the DF nodal and antinodal results.  
The normalized spectral functions, $A_k(\omega)$, appear in the inset.  Indeed, we find a non-zero density of states at the Fermi level ($\omega=0$) due primarily to thermal excitations.  
We do not observe a clear $\omega=0$ FL peak and the value of $A(\omega=0)$ for the antinodal point is $\approx 15\%$ of the nodal value an indication of the erosion of states at the Fermi level. 

\begin{figure}
\centering
    \includegraphics[width=0.95\linewidth]{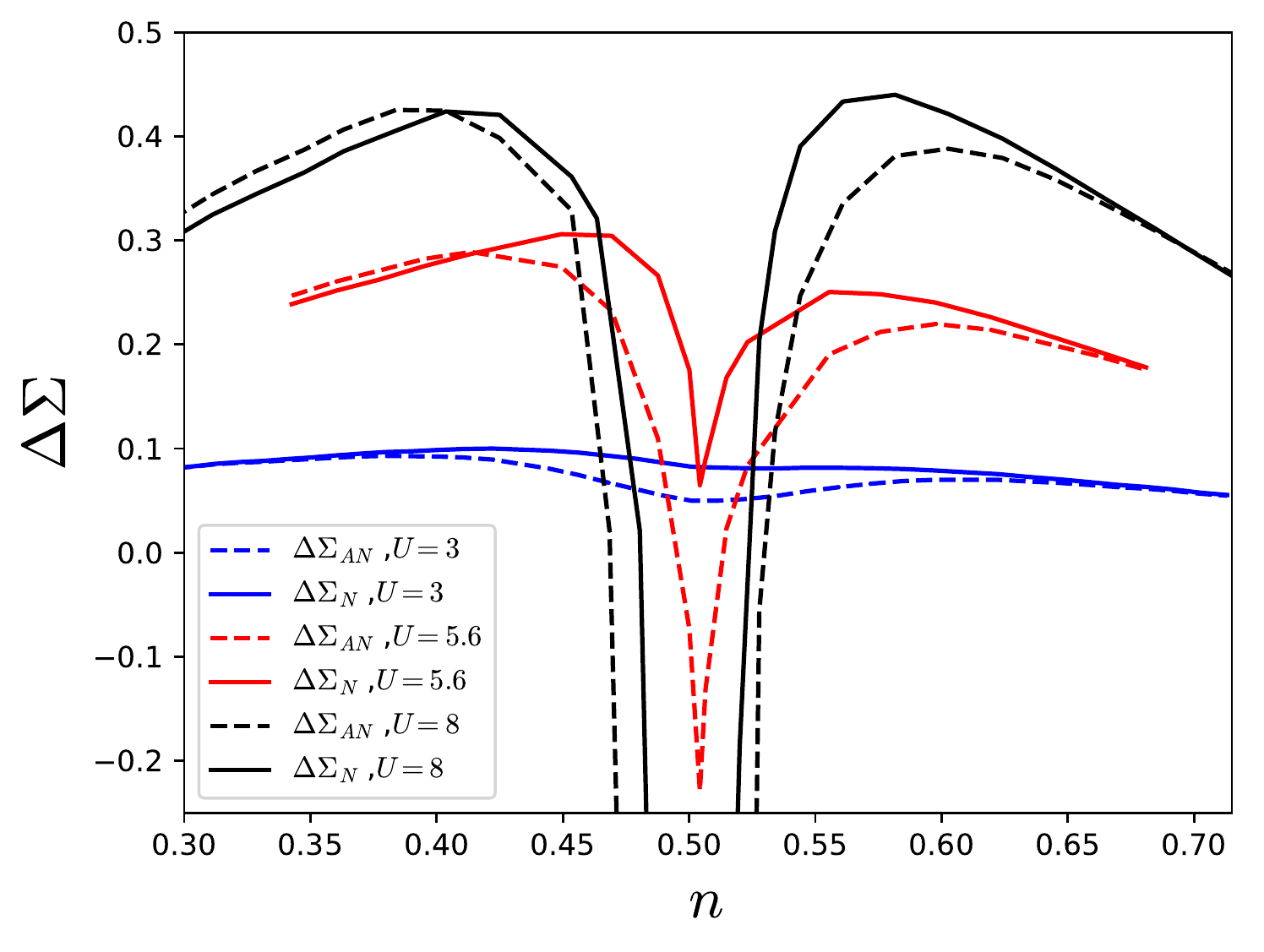}
\caption{ $\Delta \Sigma$ at the nodal and antinodal momenta as a function of density for various $U/t$ at $\beta t=5$ and $t^\prime /t=-0.3$.  \label{fig:deltasigmal}}
\end{figure}

Next we consider the $U/t$ and density, $ n $, dependence of $\Delta \Sigma$ at fixed temperature.  Results are shown in Fig.~\ref{fig:deltasigmal} where we plot the value of $\Delta \Sigma$ at the nodal and antinodal points.  At this high temperature we see that at $U/t=3$, $\Delta \Sigma$ is positive for all densities representing FL states at all momenta. 
At $U/t=5.6$ we see a region of density near half-filling where $\Delta \Sigma_{AN}<0$ while $\Delta\Sigma_N$ is always positive representing a mixed FL/nFL momentum separation near half-filling.   For $U/t=8$, both the nodal and antinodal points show nFL behaviour over a range of densities (wider for the AN point) becoming positive with either electron or hole doping away from half-filling.\cite{Gull10_clustercompare,gull:2009}
These observations at high temperature of a crossover with interaction strength are in-line with the long established cDMFT phase diagram.\cite{park:2008} What is unclear is the physical origin of the nFL behaviour, if it is a first order Mott transition or is caused by AF spin fluctuations.

\begin{figure}
\centering
    \includegraphics[width=0.9\linewidth]{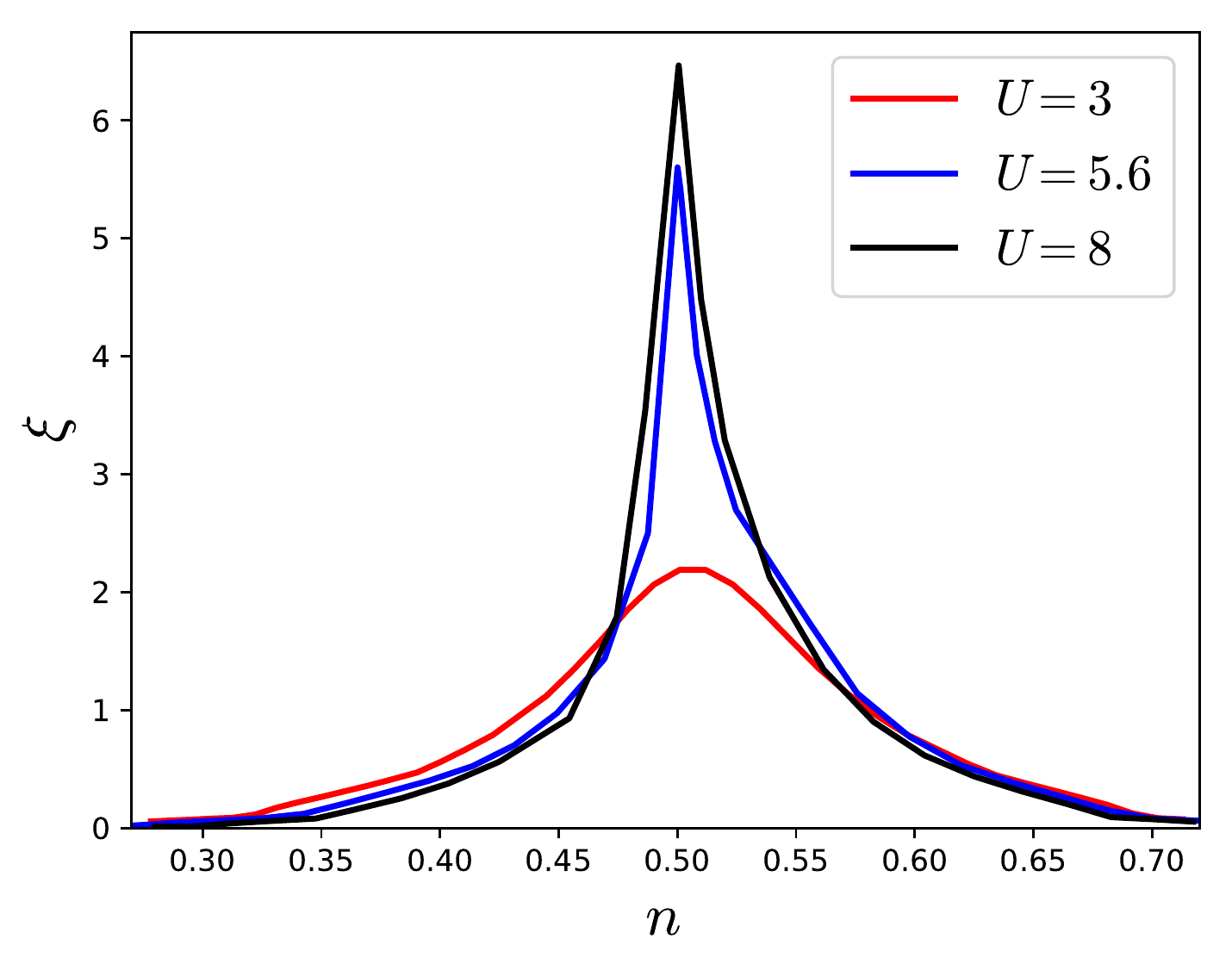}
        \caption{\label{fig:corlengthdoping}Density dependence of the anti-ferromagnetic correlation length, $\xi$, for several interaction strengths $U$ at $T/t=0.2$. obtained from a fit of $\chi_{sp}(q_x,q_y,\Omega=0)$ with the function $f(q_x,\xi)=A/((\mathbf{q}-(\pi,\pi))^2+\xi^{-2}) +c $, averaged over the $q_x=q_y$ and $(q_x,q_y)=(q_x,\pi)$ directions.\cite{rohringer:2016,gukelberger:2017}}
\end{figure}

To this end we consider the two-particle spin susceptibility\cite{hafermann:2008,chen:2017} from which we extract the correlation length as the half-width at half-max of $\chi_{sp}(q_x,q_y,i\Omega=0)$ near $(q_x,q_y)=(\pi,\pi)$.  We present this as a function of doping in Fig.~\ref{fig:corlengthdoping}.  By comparison to Fig.~\ref{fig:deltasigmal} we see the qualitative connection between the nFL and FL behavior in the self energy as a function of doping to the increase in spin-correlation length, $\xi$.  
At this high temperature for $U/t=3$ the value of $\xi \lesssim 2$ lattice sites is quite small and comparing with Fig.~\ref{fig:deltasigmal} we see that $\Delta \Sigma$ is positive for all densities.  At $U/t=5.6$ and 8, $\xi$ has a much larger value for a range of dopings near $n=1$.  It seems that the increase of $\xi$ with doping coincides with a reduction of $\Delta \Sigma$ ultimately resulting in a change of sign.  Interestingly, at these high temperatures with modest $t^\prime$, the value of $\Delta \Sigma$ is not substantially distinct between hole or electron doping ($n<1$ and $n>1$ respectively).  The antinodal point shows tendency towards nFL behavior for both for electron and hole doped cases. 
One might conclude from this that it is the tendency towards antiferromagnetic behavior that is driving the FL/nFL crossover.  At this stage, the connection is speculative and it is precisely this speculative conclusion we wish to resolve in the following section.

\begin{figure}
\centering
    \includegraphics[width=0.68\linewidth]{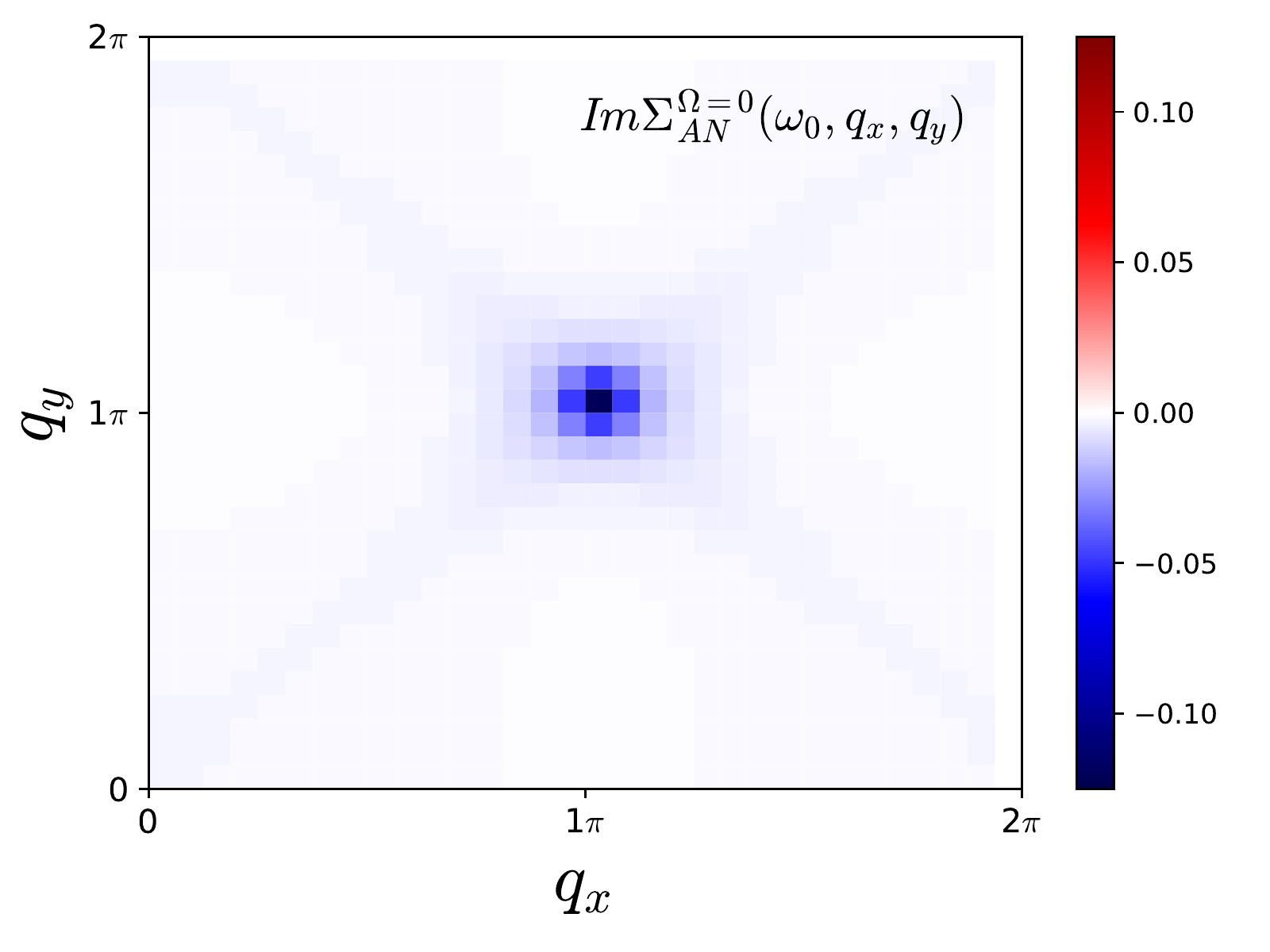}
    \includegraphics[width=0.68\linewidth]{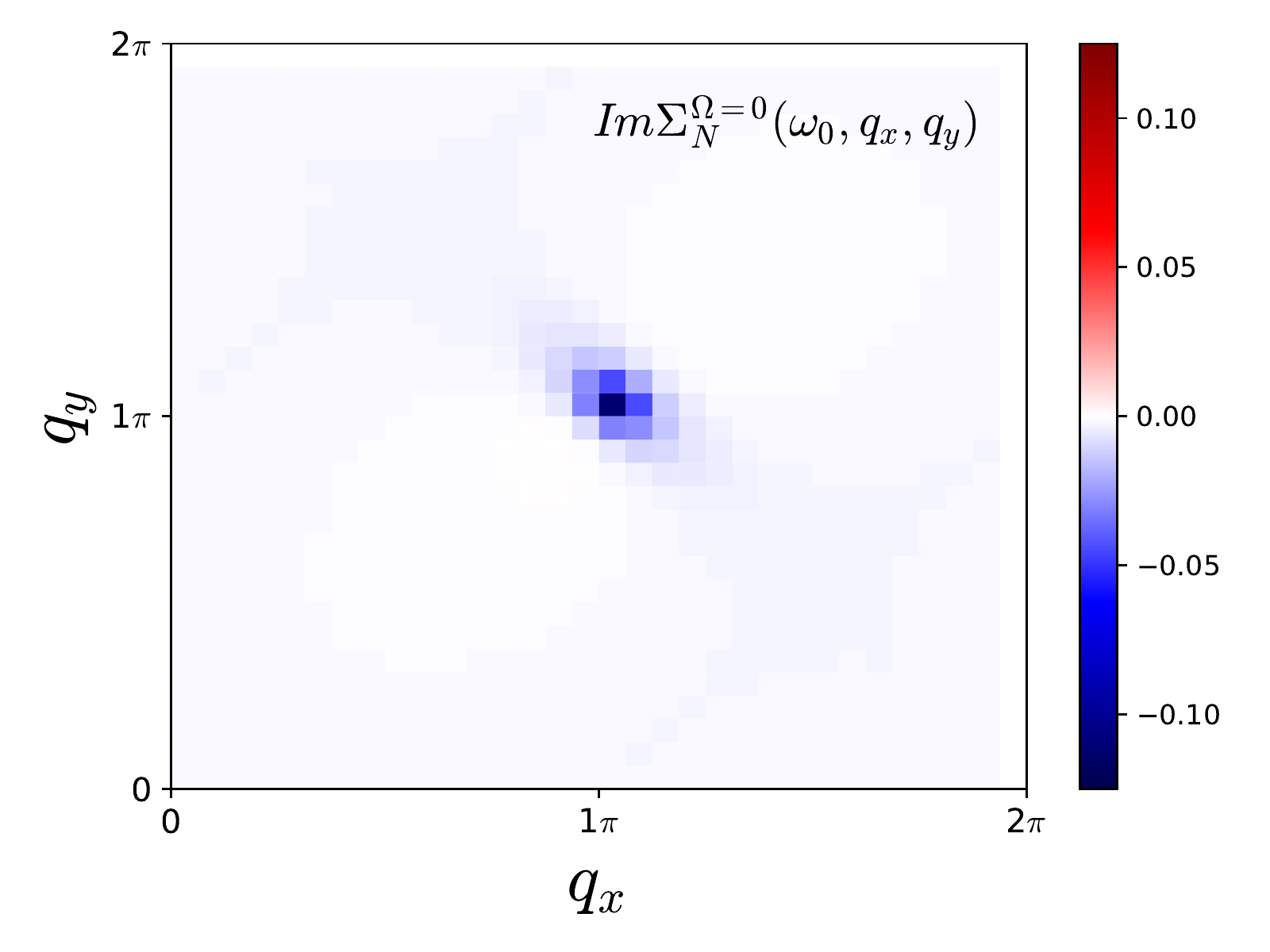}
\caption{ Imaginary part of the self energy for the antinodal (Top) and nodal (Bottom) momenta decomposed into the scattering $q=(q_x,q_y)$ contributions for the zeroth bosonic Matsubara freq for $U/t=5.6$, $\beta t =5$, $t^\prime /t=-0.3$, and $\mu=0$.\label{fig:qxqyomega0} }
\end{figure} 

\subsection{Fluctuation Diagnostics}
Up to this point we have shown only single or two particle properties.  Next we bridge the gap using Eqns.~(\ref{eqn:fluct})$\to$(\ref{eqn:sumnotation}) and examine the decomposition of the single particle $\Delta\Sigma$ into the scattering momenta and frequency channels.
We begin our discussion by illustrating  the $q-$vector deconstruction in Fig.~\ref{fig:qxqyomega0} where we show the fully deconstructed self energy contributions to the zeroeth fermionic Matsubara freq $i\omega_0$ at zero bosonic frequency as a function of the $q-$vector components. 
Similar to the results of Refs.~\onlinecite{gunnarsson:2015,wu:2017} we see a strong peak at $q=(\pi,\pi)$ for both the N and AN $k-$vectors, and weaker contributions at other $q-$vectors.  
We see that essentially all $q$-vector contributions to the self energy are negative, consistent with other work.\cite{andergassen:2019,gunnarsson:2015}
In order for $\Delta \Sigma$ to change sign there must also be positive contributions.

\begin{figure}
\centering
       \includegraphics[width=0.95\linewidth]{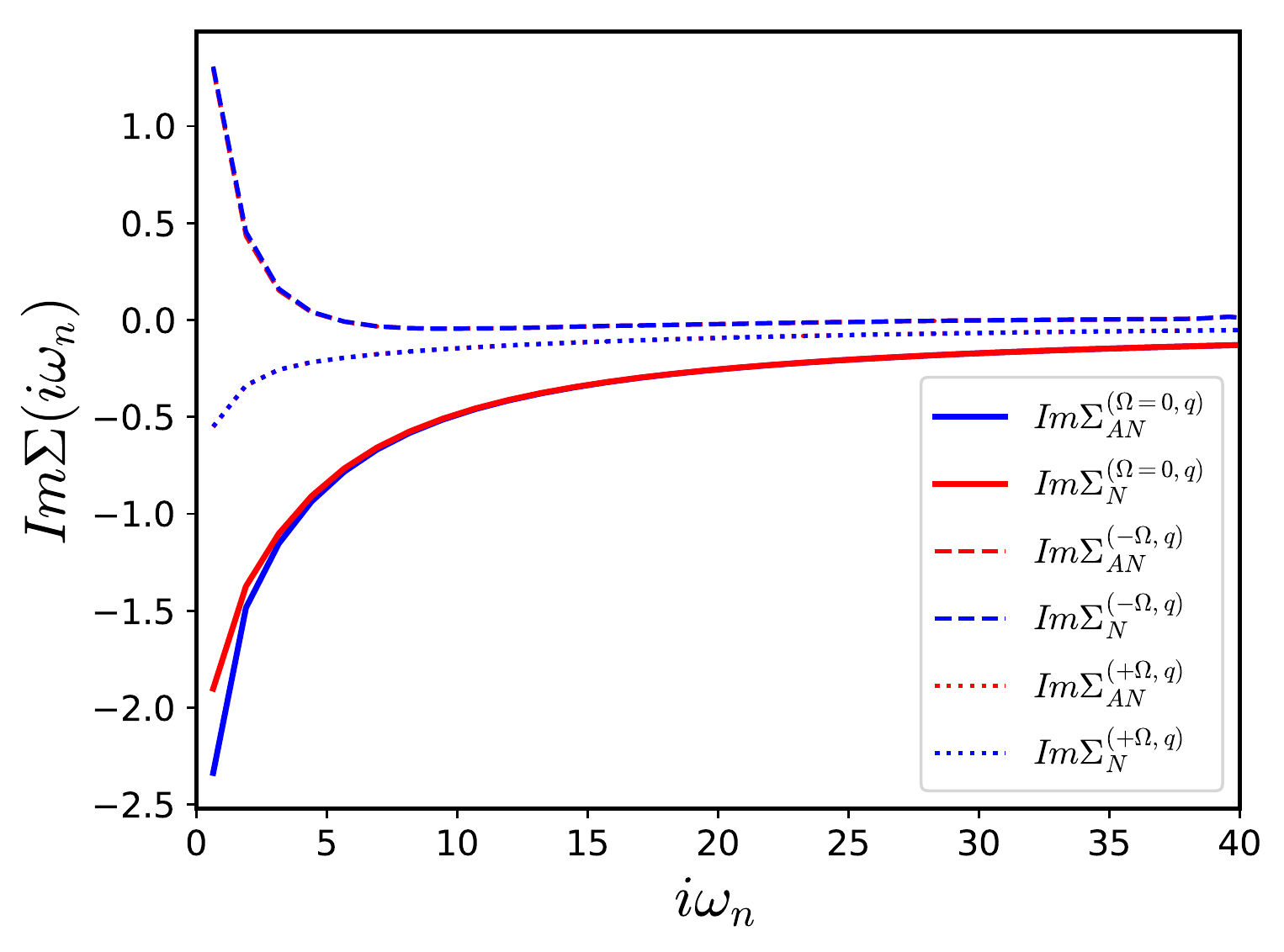}
        \caption{\label{fig:posnegbose} The imaginary part of the single particle self energy from Fig.~\ref{fig:selfedmft} deconstructed into its  $+\Omega$ and $-\Omega$ contributions and the $\Omega=0$ contribution at $U/t=5.6$, $\beta t =5$, $t^\prime /t=-0.3$, and $\mu=0$. }
\end{figure}

With this in mind we examine the behaviour of the self energy decomposed into bosonic frequency contributions such that the total self energy is just the sum $Im\Sigma_k(\omega)=Im\Sigma_k^{(\Omega,q)}= Im\Sigma_k^{(+ \Omega,q)}+Im\Sigma_k^{(- \Omega,q)} + Im\Sigma_k^{(\Omega=0,q)}$.  Results for each component are shown in Fig.~\ref{fig:posnegbose}.   We find that the largest contributions to the self energy come from $\Omega=0$ which are negative for all $i\omega_n>0$ and also show a negative contribution to $\Delta \Sigma$.   The summations over positive and negative bosonic frequencies decay rapidly with fermionic frequency while the $\Omega=0$ contribution dominates the high frequency behavior.  We also observe that the positive bosonic contributions are primarily negative while the negative bosonic frequency contributions are primarily positive.  Interestingly, the summations over $\Omega>0$ and $\Omega<0$ show almost no momentum dependence.  All of the momentum dependence comes from $\Omega=0$ excitations which provide a slightly more negative contribution to $\Delta \Sigma_{AN}$ than for $\Delta \Sigma_N$. Further, the only positive contributions to $\Delta \Sigma$ come from the summation over $\Omega<0$.  Thus, deciding if $\Delta \Sigma$ is positive or negative rests on a subtle interplay between these three components.   This is an element not mentioned in the original fluctuation diagnostics description.\cite{gunnarsson:2015} 

\begin{figure}
\centering
    \includegraphics[width=\linewidth]{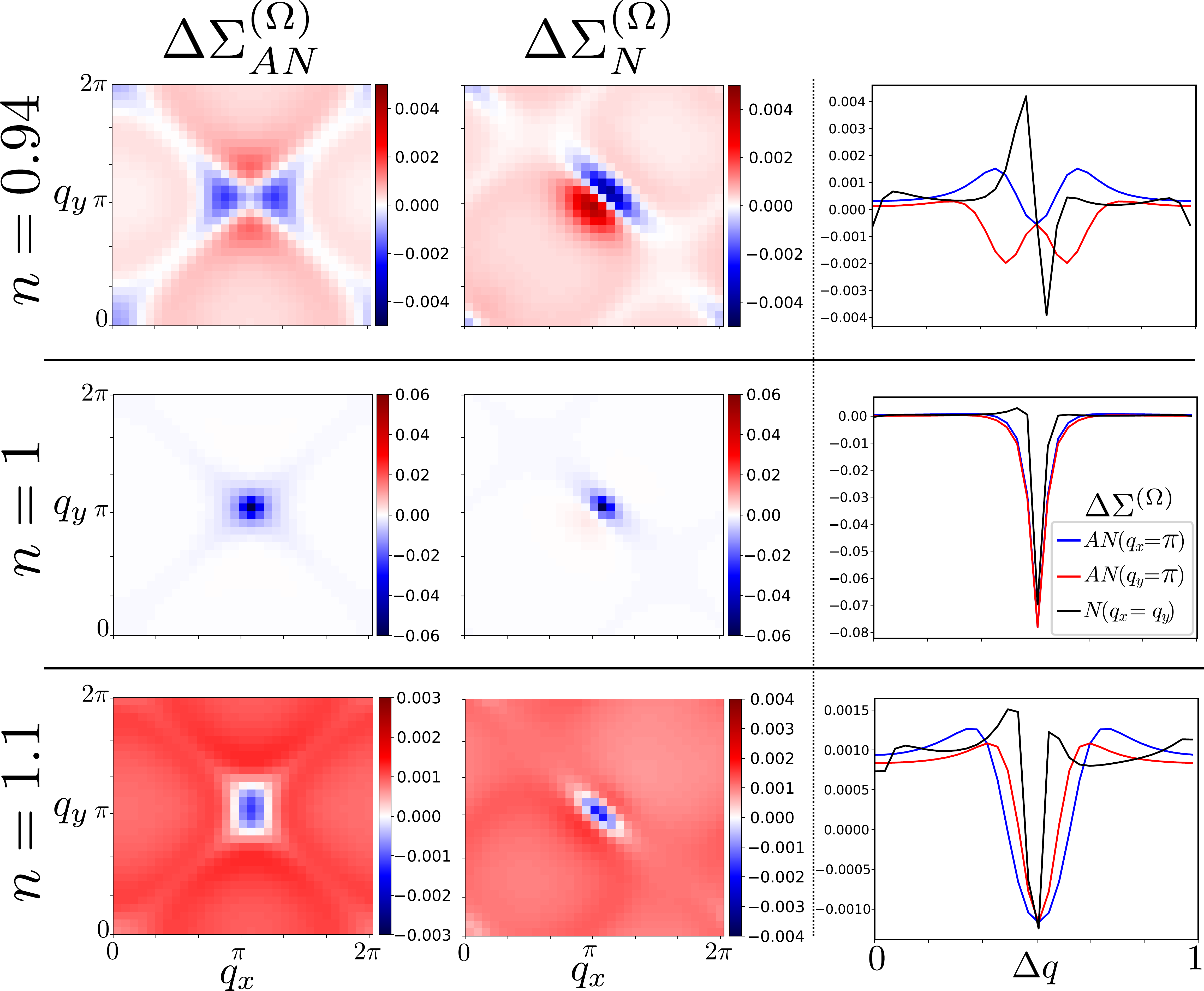}
\caption{\label{fig:fluct_DMFT_muvary} Color Plots: $\Delta \Sigma^{(\Omega)}(q_x,q_y)$ for nodal and anti- nodal momenta at $U/t= 5.6$ for $n=0.94$ (top row), 1 (middle) and 1.1 (bottom row) corresponding to chemical potentials of $\mu/t = -0.8, 0, 1.4 $ respectively.  Right handle column shows cuts of $\Delta \Sigma^{(\Omega)}(q_x,q_y)$ for: (blue) a cut in the antinodal result along the path from $(0,\pi) \to (2\pi,\pi)$, (red) a cut in the antinodal result along the path from $(\pi,0) \to (\pi,2\pi)$, and (black) a cut in the nodal result along the path from $(0,0)\to(2\pi,2\pi)$. Each path is plotted by its length in the x-axis, normalized by the total length of the cut. }
\end{figure}

We explore also the doping dependence wherein we select a specific range of densities for which expect $\Delta \Sigma$ to switch sign. 
Results for $n=0.94, 1.0,$ and 1.1 are displayed in Fig.~\ref{fig:fluct_DMFT_muvary} for $\Delta \Sigma^{(\Omega)}(q_x,q_y)$ which includes the total bosonic contributions at each $q_x$ and $q_y$ value.  
This allows us to see specifically which $q$-vectors give FL ($\Delta \Sigma^{(\Omega)}>0$) contributions or nFL  ($\Delta \Sigma^{(\Omega)}<0$) contributions.  We focus on $U/t=5.6$ where nodal and antinodal differences occur. 
To assist with analyzing the color plots, the right hand column of Fig.~\ref{fig:fluct_DMFT_muvary} includes high symmetry cuts through the Brillouin zone.

Starting first with the top row of Fig.~\ref{fig:fluct_DMFT_muvary} we see that both the antinodal and nodal frames (left and middle columns respectively) show complicated structure including both positive and negative contributions.
To assist with the discussion we also provide horizontal, vertical, and diagonal cuts through the datasets in the right hand column.
From Fig.~\ref{fig:deltasigmal} we know that the total antinodal $\Delta \Sigma_{AN}<0$ while the total nodal $\Delta \Sigma_{N} >0$ when these results are summed over $q_x$ and $q_y$. 
For the anti-nodal frame most $q$-vectors provide positive contributions, but are very weak, with strong positive contributions above and below $(\pi,\pi)$.  These positive contributions are just slightly overpowered by the strong negative contributions to the left and right of $(\pi,\pi)$.
In the nodal case again the majority of $q$-vectors give weak positive contribution and we observe both strong negative and positive peaks.  In this case the strong positive feature near $(\pi,\pi)$ overpowers the strong negative feature giving an overall $\Delta \Sigma_N >0$.

The situations for $n=1$ and $n=1.1$ are less subtle.  At half filling all $q$-vectors give negative contributions, with the strongest feature near $q=(\pi,\pi)$. 
Interestingly, at $n=1.1$  the $(\pi,\pi)$ feature remains strong and negative but is now overwhelmed by the background of all other $q$-vectors which give positive contributions resulting in both the nodal and antinodal columns having a total $\Delta\Sigma>0$ as we know is the case from Fig.~\ref{fig:deltasigmal}.
This result supports the often noted idea that the physics of hole and electron doped systems are quite distinct with finite $t^\prime$. 
Finally, from the original fluctuation diagnostics work\cite{gunnarsson:2015} we understand that the observation of the broad background for $n=1.1$ might suggest that the spin channel is not the best (most compact) basis for describing the electron doped system.

\subsection{Temperature Dependence at Weak-Coupling}

\begin{figure}
\centering
    \includegraphics[width=0.95\linewidth]{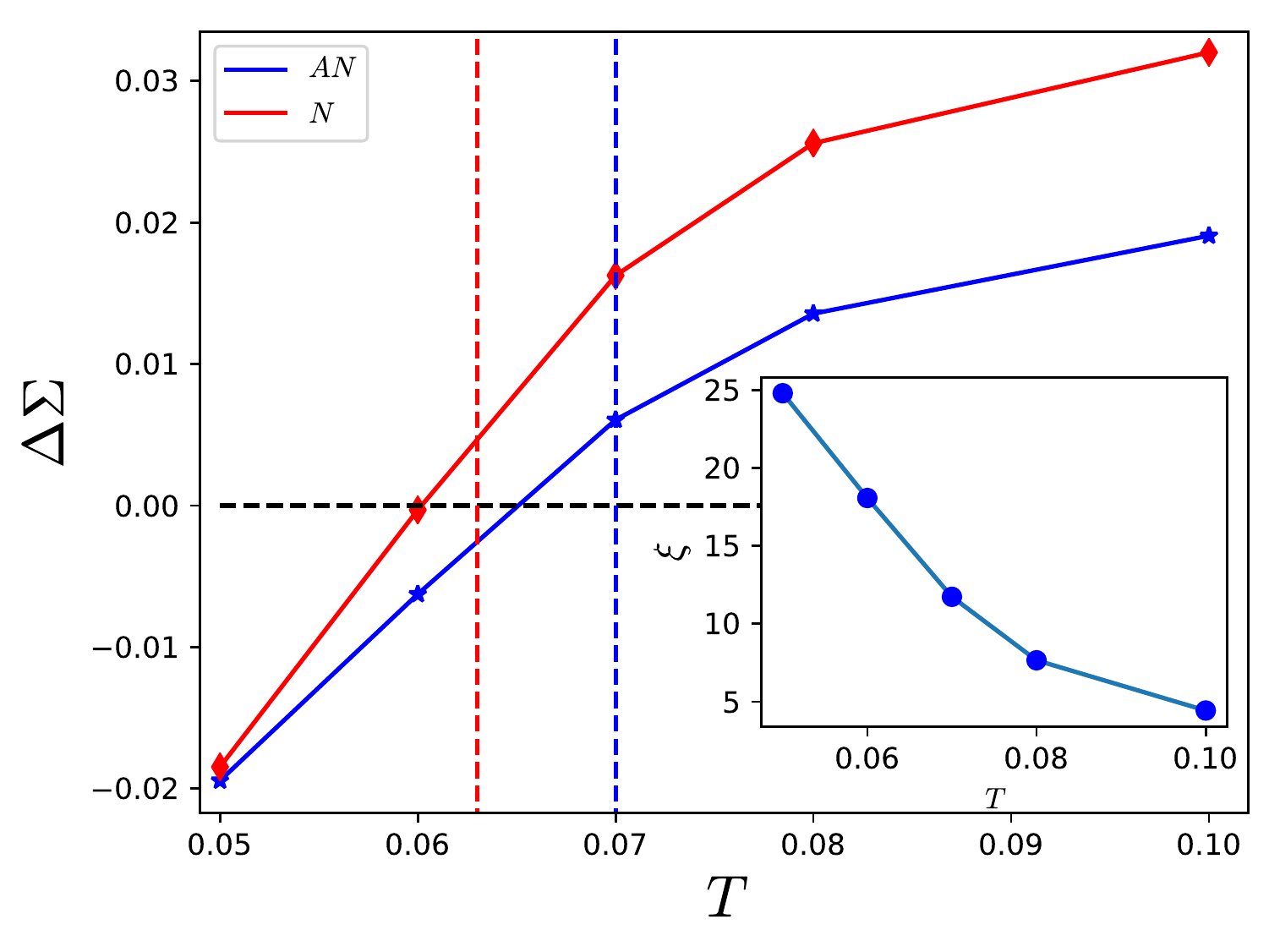}
\caption{\label{fig:lowt} $\Delta \Sigma$ vs temperature for $U/t$=2, $t^\prime=0$, $\mu=0$ for nodal and antinodal momenta.  Vertical lines are $\Sigma - DDMC$ crossover points identified in Ref.~\onlinecite{simkovic:2019}.  Inset: Correlation length scale $\xi$ as a function of temperature in units of hopping, $T/t$. }
\end{figure} 

\begin{figure}
\centering
        \includegraphics[width=\linewidth]{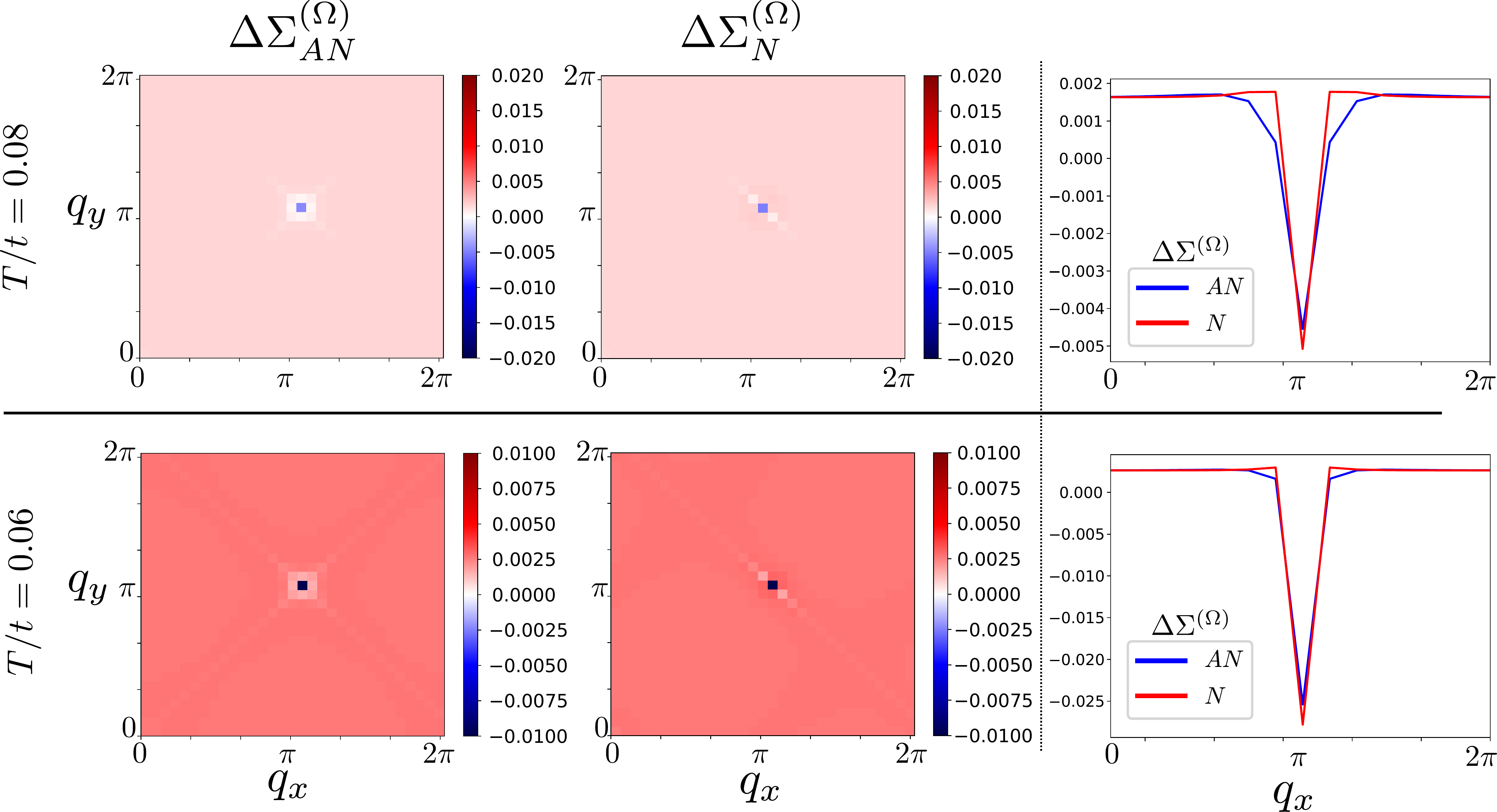}
        \caption{ \label{fig:lowt2}$\Delta \Sigma_{N/AN}^{\Omega} (q_x,q_y)$  values for $U/t=2$, $t^{\prime}=0$ $\mu=0$ for temperatures $T/t=0.08$ (top row) and $T/t=0.06$ (bottom row). Right handle column shows cuts of $\Delta \Sigma^{(\Omega)}(q_x,q_y)$ for $q_y=\pi$ as a function of $q_x$.
        }
    \end{figure}

In this section we target the half-filled, particle-hole symmetric case at weak coupling.  We examine the $\beta t$ dependence for parameters $U/t=2$, $t^\prime=0$, and $\mu=0$ for which results are shown in Fig.~\ref{fig:lowt} for $\beta t=10 \to 20$. 
At $\beta t=10$ $\Delta \Sigma$ is positive for both nodal and antinodal momenta and its value decreases with decreasing temperature.
 We find that $\Delta\Sigma$ changes sign at sufficiently low temperature, $T/t\approx0.06$ for $\Delta \Sigma_N$ and at $T/t\approx0.065$ for $\Delta \Sigma_{AN}$ which we estimate from the linear interpolation of Fig.~\ref{fig:lowt}.  
 These crossovers from FL to nFL physics are in qualitative agreement, though slightly lower in temperature, with exact results from $\Sigma-DDMC$\cite{simkovic:2019} represented by vertical lines at $T/t=0.063$ and $0.07$ for nodal and antinodal points respectively (determined from the analytic fit in Ref.~\onlinecite{simkovic:2019}).  The deviation between the exact $\Sigma-DDMC$ results and our DF calculations appears fundamental to the DF approximation and is not improved when increasing our $k$-space grid resolution, frequency grids or convergence criteria.
We extract the correlation length scale from $\chi_{sp}(q_x,q_y,\Omega=0)$, plotted in the inset over the same temperature range, and note that on approach to the nodal crossover the spin correlation length is very large $\xi\approx 15$.  This is similar to cases at  $U/t \approx 5\to 8$ at higher temperature where the shape of $\xi$ with doping is similar to that of $\Delta \Sigma$ with doping.  Overall, these results support the conclusions of Ref.~\onlinecite{simkovic:2019} that this nFL crossover is driven by AFM fluctuations. 

As was done for doping dependent results, we show the decomposed quantities in Fig.~\ref{fig:lowt2} at $T/t=0.06$ and $T/t=0.08$ just above and below the $\Delta \Sigma=0$ line.  Interestingly, in both cases the majority of the q-vectors provide positive contributions and these actually increase as temperature decreases.  We see only a strong, sharp negative contribution precisely at $q=(\pi,\pi)$.  This is in extreme contrast to the earlier examples where both doping and finite $t^\prime$ suppress and broaden the $(\pi,\pi)$ feature.  Indeed, for the particle-hole symmetric case, all of the negative, nFL, contributions come from $(\pi,\pi)$ AFM fluctuations.  This result unequivocally shows that it is $(\pi,\pi)$ AF fluctuations that are responsible for the FL to nFL crossover observed in Ref.~\onlinecite{simkovic:2019}.

\section{Concluding Remarks \label{sec:conclusions}} 

 We examined the doping dependence at high temperature of $\Delta \Sigma$ and of the correlation length scale and found that $\Delta \Sigma$ mimics the antiferromagnetic correlation length $\xi$ both with doping at fixed temperature and with temperature at fixed doping.  We observe that the tendency towards nFL behavior is nearly equivalent for both densities above and below half-filling even at high temperature.  Such an observation may have important implications for studies of electron doped high-Tc materials where pseudogaps have been observed.\cite{yamada:2005,damascelli:2018}
To strengthen this connection we performed a fluctuation diagnostic decomposition of the self energy $\Delta \Sigma$ to show that the negative (nFL) contributions are primarily near $(\pi,\pi)$ for doping close to half-filled which suggests they are AFM in origin.  Contributions for $q$-vectors away from $(\pi,\pi)$ are expected for doped cases, and within the framework of fluctuation diagnostics, the observation of sharp features suggests that spin excitations remain a good basis for representing the fundamental excitations responsible for the nFL behavior for weakly doped systems in the single band model.\cite{gunnarsson:2015}
 We examine as well the weak coupling regime $U/t=2$ and find a magnetically driven FL/nFL crossover  at finite temperature in agreement with Ref.~\onlinecite{simkovic:2019}.
The result is a non-trivial finite correlation length that non-the-less leads to non-Fermi-liquid behaviour.   The physics of the interaction driven Mott-transition underneath this nFL/quasi-AFM state remains unknown.
One possibility is that the quasi-AFM state gaps the system and preempts the Mott transition from occurring.  However, it might be that both physical processes occur leading to a Mott-transition from a quasi-AFM insulating phase to a Mott insulating phase, or that both effects coincide. 
With this in mind, further exploration to show the role of other metrics of the 1st order Mott transition might be valuable.  For example, the hysteresis in the double occupancy or Green's function, $-\beta G(\tau=\beta/2)$, for which recent work suggests that small cDMFT clusters capture correct physics.\cite{vanloon:2018:hysteresis,park:2008}   
Finally, gaining a physical intuition into which particular $q-$vectors are responsible for metallic or insulating behaviour might allow one to construct simple effective models of complex systems.  Further study to remove the approximations inherent with the self-consistent ladder DF method may also be fruitful, such as including higher order vertex functions when constructing the dual self energy which appear to have a small but non-zero effect on $\Delta \Sigma$.\cite{ribic:2017}

\section{Acknowledgments}
We thank Markus Wallerberger for proposing fluctuation diagnostics from the dual fermion method.
JPFL acknowledges the support of the Natural Sciences and Engineering Research Council of Canada (NSERC), RGPIN-2017-04253 and from the Simons collaboration on the many-electron problem. 
Computational resources were provided by ACENET and Compute Canada.

\bibliographystyle{apsrev4-1}
\bibliography{refs.bib}

\end{document}